       \documentclass[twocolumn,amsmath,amssymb,10pt,superscriptaddress,a4paper,letterpaper,fleqn]{revtex4-1}
       \usepackage{amssymb}
       \usepackage{epsfig}
       \usepackage{graphicx}
       \usepackage{dcolumn}
       \usepackage{array}
       \usepackage{bm}
       \usepackage{fancyheadings}
       \usepackage{longtable}
       \usepackage{multirow}
       \usepackage{float}
       \usepackage{afterpage}
       \usepackage{color}
       \usepackage{txfonts}
       
       \setlongtables

       \def\EXFOR{${\rm EXFOR}$ }
       \def\etal{\textit{et al.}}
       \def\ie{\textit{i.e.,~}}
       \def\eg{\textit{e.g.,~}}
       
\begin{document}
       \setcounter{page}{1}
       
       
       \title{
       \qquad \\ \qquad \\ \qquad \\  \qquad \\  \qquad \\ \qquad \\
       Toward More Complete and Accurate Experimental Nuclear Reaction Data
       Library (EXFOR) -
       International Collaboration Between Nuclear Reaction Data Centres (NRDC)
       }
       
       \author{N.~Otuka}
       \email[Corresponding author: ]{n.otsuka@iaea.org}
       \affiliation{
       Nuclear Data Section (NDS), International Atomic Energy Agency,
       A-1400 Wien, Austria}
       
       \author{E.~Dupont}
       \affiliation{
       OECD Nuclear Energy Agency Data Bank (NEA DB),
       F-92130 Issy-les-Moulineaux, France}
       
       \author{V.~Semkova}
       \affiliation{
       Nuclear Data Section (NDS), International Atomic Energy Agency,
       A-1400 Wien, Austria}
       
       \author{B.~Pritychenko} 
       \affiliation{
       National Nuclear Data Center (NNDC), Brookhaven National Laboratory,
       Upton, NY 11973, USA} 
       
       \author{A.I.~Blokhin} 
       \affiliation{
       Nuclear Data Centre (CJD), Institute for Physics and Power Engineering,
       249033 Obninsk, Russia} 
       
       \author{M.~Aikawa} 
       \affiliation{
       Nuclear Reaction Data Centre (JCPRG), Hokkaido University,
       Sapporo 060-0810, Japan} 
       
       \author{S.~Babykina} 
       \affiliation{
       Centre for Nuclear Structure and Reaction Data (CAJaD),
       Kurchatov Institute, 123182 Moscow, Russia} 
       
       \author{M.~Bossant}
       \affiliation{
       OECD Nuclear Energy Agency Data Bank (NEA DB),
       F-92130 Issy-les-Moulineaux, France}
       
       \author{Chen~Guochang}
       \affiliation{
       China Nuclear Data Centre (CNDC), China Institute of Atomic Energy,
       Beijing 102413, China} 
       
       \author{S.~Dunaeva}
       \affiliation{
       Centre of Nuclear Physics Data (CNPD), All-Russian Research Institute of Experimental Physics (VNIIEF),
       607190 Sarov, Russia} 
       
       \author{R.A.~Forrest} 
       \affiliation{
       Nuclear Data Section (NDS), International Atomic Energy Agency,
       A-1400 Wien, Austria}
       
       \author{T.~Fukahori} 
       \affiliation{
       Nuclear Data Center, Japan Atomic Energy Agency (JAEA),
       Tokai-mura, Naka-gun, Ibaraki 319-1195, Japan} 
       
       \author{N.~Furutachi} 
       \affiliation{
       Nuclear Reaction Data Centre (JCPRG), Hokkaido University,
       Sapporo 060-0810, Japan} 
       
       \author{S.~Ganesan} 
       \affiliation{
       Bhabha Atomic Research Centre, Mumbai 400085, India}
       
       \author{Ge~Zhigang} 
       \affiliation{
       China Nuclear Data Centre (CNDC), China Institute of Atomic Energy,
       Beijing 102413, China} 
       
       \author{O.O.~Gritzay} 
       \affiliation{
       Ukrainian Nuclear Data Centre (UkrNDC), Institute for Nuclear Research,
       03680 Kyiv, Ukraine} 
       
       \author{M.~Herman} 
       \affiliation{
       National Nuclear Data Center (NNDC), Brookhaven National Laboratory,
       Upton, NY 11973, USA} 
       
       \author{S.~Hlava\v c} 
       \affiliation{
       Department of Nuclear Physics, Institute of Physics, Slovak Academy of Science,
       845 11 Bratislava, Slovakia} 
       
       \author{K.~Kat\=o} 
       \affiliation{
       Nuclear Reaction Data Centre (JCPRG), Hokkaido University,
       Sapporo 060-0810, Japan} 
       
       \author{B.~Lalremruata} 
       \affiliation{
       Department of Physics, Mizoram University, Aizawl 796004, India} 
       
       \author{Y.O.~Lee} 
       \affiliation{
       Korea Nuclear Data Center (KNDC),
       Korea Atomic Energy Research Institute,
       Daejeon 305-600, Republic of Korea}
       
       \author{A.~Makinaga} 
       \affiliation{
       Nuclear Reaction Data Centre (JCPRG), Hokkaido University,
       Sapporo 060-0810, Japan} 
       
       \author{K.~Matsumoto}
       \affiliation{
       OECD Nuclear Energy Agency Data Bank (NEA DB),
       F-92130 Issy-les-Moulineaux, France}
       
       \author{M.~Mikhaylyukova} 
       \affiliation{
       Nuclear Data Centre (CJD), Institute for Physics and Power Engineering,
       249033 Obninsk, Russia} 
       
       \author{G.~Pikulina}
       \affiliation{
       Centre of Nuclear Physics Data (CNPD), All-Russian Research Institute of Experimental Physics (VNIIEF),
       607190 Sarov, Russia} 
       
       \author{V.G.~Pronyaev} 
       \affiliation{
       Nuclear Data Centre (CJD), Institute for Physics and Power Engineering,
       249033 Obninsk, Russia} 
       
       \author{A.~Saxena} 
       \affiliation{
       Bhabha Atomic Research Centre,  Mumbai 400085, India}
       
       \author{O.~Schwerer} 
       \affiliation{
       Under contract with National Nuclear Data Center, Brookhaven National Laboratory,
       Upton, NY 11973, USA} 
       
       \author{S.P.~Simakov} 
       \affiliation{
       Nuclear Data Section (NDS), International Atomic Energy Agency,
       A-1400 Wien, Austria}
       
       \author{N.~Soppera}
       \affiliation{
       OECD Nuclear Energy Agency Data Bank (NEA DB),
       F-92130 Issy-les-Moulineaux, France}
       
       \author{R.~Suzuki}
       \affiliation{
       Nuclear Reaction Data Centre (JCPRG), Hokkaido University,
       Sapporo 060-0810, Japan} 
       
       \author{S.~Tak\'acs} 
       \affiliation{
       Cyclotron Application Department,
       Institute of Nuclear Research (ATOMKI), H-4001 Debrecen, Hungary} 
       
       \author{Tao~Xi} 
       \affiliation{
       China Nuclear Data Centre (CNDC), China Institute of Atomic Energy,
       Beijing 102413, China} 
       
       \author{S.~Taova}
       \affiliation{
       Centre of Nuclear Physics Data (CNPD), All-Russian Research Institute of Experimental Physics (VNIIEF),
       607190 Sarov, Russia} 
       
       \author{F.~T\'ark\'anyi} 
       \affiliation{
       Cyclotron Application Department,
       Institute of Nuclear Research (ATOMKI), H-4001 Debrecen, Hungary} 
       
       \author{V.V.~Varlamov} 
       \affiliation{
       Centre for Photonuclear Experiments Data (CDFE),
       Institute of Nuclear Physics, Moscow State University,
       119234 Moscow, Russia} 

       \author{Wang~Jimin} 
       \affiliation{
       China Nuclear Data Centre (CNDC), China Institute of Atomic Energy,
       Beijing 102413, China} 
       
       \author{S.C.~Yang} 
       \affiliation{
       Korea Nuclear Data Center (KNDC),
       Korea Atomic Energy Research Institute,
       Daejeon 305-600, Republic of Korea}
       
       \author{V.~Zerkin}
       \affiliation{
       Nuclear Data Section (NDS), International Atomic Energy Agency,
       A-1400 Wien, Austria}
       
       \author{Zhuang~Youxiang}
       \affiliation{
       China Nuclear Data Centre (CNDC), China Institute of Atomic Energy,
       Beijing 102413, China} 
       
       \date{\today} 
       
       \begin{abstract}
       {
       %
       The International Network of Nuclear Reaction Data Centres (NRDC)
       coordinated by the IAEA Nuclear Data Section (NDS) is successfully
       collaborating in the maintenance and development of the \EXFOR library.
       As the scope of published data expands
       (\eg to higher energy, to heavier projectile) to meet the needs from
       the frontier of sciences and applications,
       it becomes nowadays a hard and challenging task to maintain both
       completeness and accuracy of the whole \EXFOR library.
       The paper describes evolution of the library with highlights on recent
       developments.  
       }
       \end{abstract}
       \maketitle
       
       \lhead{Toward More Complete and $\dots$}
       \chead{NUCLEAR DATA SHEETS}
       \rhead{N. Otuka \textit{et al.}}
       \lfoot{}
       \rfoot{}
       \renewcommand{\footrulewidth}{0.4pt}
       
       
       \section{INTRODUCTION}
       The \EXFOR library has become the most comprehensive compilation of
       experimental nuclear reaction data.
       It contains cross sections and other nuclear reaction quantities
       induced by neutron, charged-particle and photon beams.
       Currently compilation is mandatory for all low and intermediate
       energy ($\le$ 1 GeV) neutron and light charged-particle ($A\le 12$)
       induced reaction data.
       Heavy-ion ($A \ge 13$) and photon induced reaction data are also
       additionally compiled on a voluntary basis.
       
       Currently fourteen data centers shown in Table~\ref{tab:1} are participating
       in the International Network of Nuclear Reaction Data Centres
       (NRDC)~\cite{NO10} and are collaborating
       mainly for compilation and exchange of
       experimental data by using the common Exchange Format
       (\EXFOR format)~\cite{NO11a} under the auspices of the IAEA Nuclear Data
       Section (NDS).
       \begin{table}
       \centering
       \caption{Scope of compilation and \EXFOR web retrieval service
                (ND/CPND/PhND: neutron/charged-particle/photonuclear data).}
       \label{tab:1}
       \begin{tabular}{ll} 
       \hline
       \hline
       Center &Scope and URL\\
       \hline
       ATOMKI &CPND measured in collaboration with ATOMKI\\
       CAJaD  &CPND measured in former USSR (except for Ukraine)\\
       CDFE   &PhND (coordinated with other centers)\\
              &~http://cdfe.sinp.msu.ru/exfor/\\
       CJD    &ND measured in former USSR (except Ukraine)\\
       CNDC   &ND and CPND measured in China\\
              &~http://www-nds.ciae.ac.cn/exfor/\\
       CNPD   &CPND (coordinated with other centers)\\
       JAEA   &Evaluation.  http://spes.jaea.go.jp/\\
       JCPRG  &CPND and PhND measured in Japan\\
              &~http://www.jcprg.org/exfor/\\
       KNDC   &ND, CPND and PhND measured in Korea\\
       NDPCI\protect\footnote{NDPCI: Nuclear Data Physics Centre of India (virtual center)}
              &ND, CPND and PhND measured in India\\
              &~http://www-nds.indcentre.org.in/exfor/\\
       NDS    &ND, CPND and PhND not covered by other centers\\
              &~http://www-nds.iaea.org/exfor/\\
       NEA DB &ND and CPND measured in NEA DB countries\\
              &not covered by other centers\\
              &~http://www.oecd-nea.org/janisweb/search/exfor/\\
       NNDC   &ND, CPND and PhND measured in USA and Canada\\
              &~http://www.nndc.bnl.gov/exfor/\\
       UkrNDC &ND, CPND and PhND measured in Ukraine\\
       \hline
       \hline
       \end{tabular}
       \end{table}
       Following an introduction to the current \EXFOR compilation procedure,
       this paper summarizes various recent efforts to make the contents of the
       \EXFOR library more complete and accurate.
       Readers interested in the history of the NRDC activity are guided to our
       previous report~\cite{NO11b} and references therein.
       
       \section{COMPILATION}
\begin{figure}[!htb]
       \includegraphics[width=0.95\columnwidth]{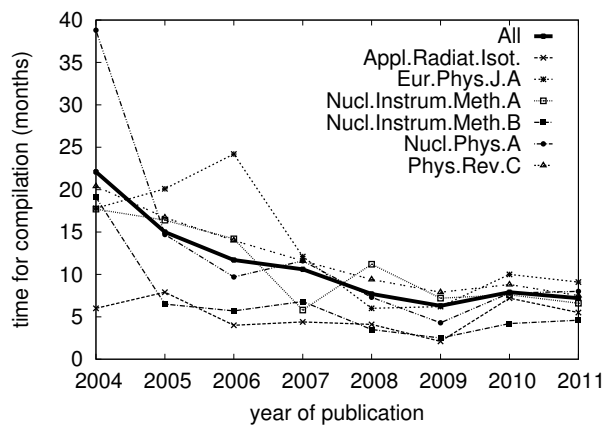}
       \caption{
       Average time for compilation of experimental data 
       (time difference between publication and inclusion to the \EXFOR Master File).
       }
       \label{fig:1}
       \end{figure}

       The first important step of data compilation is to scan literature and
       identify articles reporting experimental data for \EXFOR compilation.
       For many decades, neutron-induced reaction measurement publications were
       indexed for CINDA (Computer Index for Neutron Data) by ``CINDA readers"
       world-wide~\cite{NEA07}, and
       \EXFOR compilers could use it as the complete and independent list of
       experimental works.
       These CINDA readers are no longer available,
       and NDS is regularly scanning more than 60 journals to identify 
       articles for compilation.
       Articles identified by NDS and other data centers are registered to an
       internal database for assignment of an \EXFOR entry number by the
       responsible center
       (\eg NNDC for data measured in USA and Canada).
       Progress in compilation and distribution of compilation responsibility
       are periodically reviewed and discussed in annual NRDC meetings.
       Figure~\ref{fig:1} shows the average time for \EXFOR compilation
       (time difference between publication and inclusion to the \EXFOR Master File)
       for articles that must be compiled from
       six major journals.
       We observe that it takes 5 to 10 months on average to release
       an entry to \EXFOR users since publication.
       
       A set of new and revised \EXFOR entries is assembled to a ``tape" by
       the originating center, and transmitted to other centers 
       (preliminary transmission).
       The originating center waits for comments from other centers
       for one month at minimum,
       and transmits again a corrected tape to other centers 
       (final transmission).
       Since 2005, a complete set of the latest \EXFOR entries is maintained
       by NDS as the \EXFOR Master File and database.
       It is updated on a
       monthly basis, and its contents are available at the NDS \EXFOR web
       retrieval service~\cite{VZ08}.
       Other data centers providing their own \EXFOR retrieval services are also
       encouraged to adopt the \EXFOR Master File in order to provide the same
       contents to users.
       The newly released \EXFOR data sets are also indexed in
       the ``\EXFOR News" by NDS, and distributed to data centers as well as
       individual subscribers.
       
\begin{figure}[!htb]
       \includegraphics[width=0.95\columnwidth]{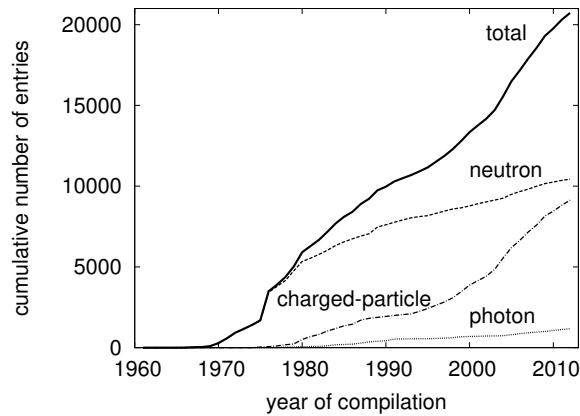}
       \caption{
       Cumulative number of \EXFOR entries (experimental works) created in each year of
       compilation.
       }
       \label{fig:2}
       \end{figure}

       Figure~\ref{fig:2} shows time evolution of the number of \EXFOR entries.
       Only neutron-induced reaction data were compiled at the beginning of
       data exchange, while compilation of charged-particle and photon
       induced reaction data was started in the middle of 1970s.
       Now the contents of neutron and charged-particle induced reaction
       data in the \EXFOR library are comparable, and more than 20000 experimental
       works are accumulated in the library.
       
       Some data centers are developing compilation tools
       (\eg editors, digitizers).
       For example, an editor developed by CNPD (\EXFOR Editor) is used by
       \EXFOR compilers for input of information in the \EXFOR format.
       Also a Java based digitizer developed by JCPRG (GSYS)~\cite{RS13}
       is used for digitization of figure images
       to extract numerical data not available from experimentalists.
       In order to utilize these compilation tools,
       NDS periodically organises workshops for \EXFOR compilers in Vienna.
       Similar workshops are also organised at regional and country levels.
       For example, four Asian data centers (CNDC, JCPRG, KNDC, NDPCI) organised
       three workshops (2010 in Sapporo, 2011 in Beijing, 2012 in Pohang) 
       to stimulate \EXFOR compilation and other nuclear reaction
       database developments.
       The Indian center (${\rm NDPCI}$) also organises \EXFOR compilation
       workshops regularly (2006 and 2007 in Mumbai, 2009 in Jaipur, 2011 in Chandigarh,
       2013 in Varanasi), and many experimental data measured in India have been compiled
       by the participants from Indian universities and institutes.
       
       \section{COMPLETENESS}
       The \EXFOR library is expected to be complete for low and intermediate
       energy neutron and light charged-particle induced reaction
       data.
       However, the coverage of light charged-particle induced reaction data
       (especially differential cross sections) is not as good as that of
       neutron induced reaction data because compilation of charged-particle
       induced reaction data was started later.

       Some examples of recent attempts to improve the coverage of the \EXFOR
       library are summarized below with the number of articles missed in \EXFOR
       in parentheses:
       \begin{itemize}
       \item[1]{\bf Neutron source spectra (30):}
       Data reporting neutron source spectra
        (\eg neutron spectra from $^9$Be+d).
       The compilation rules were also discussed in the IAEA Consultant Meeting on
       ``Neutron Source Spectra for \EXFOR"~\cite{SS11a}.
       Compilation is on-going.
       \item[2]{\bf Therapeutic radioisotope production cross sections (40):}
       Data identified within the IAEA CRP on ``Nuclear Data for Production of
       Therapeutic Radionuclides"~\cite{SMQ11}.
       All articles were compiled by 2008.
       \item[3]{\bf Isotope production cross sections (300):}
       Data for light-charged particle (p, d, t, $^{3}$He,
       $\alpha$) induced isotope production in Landolt-B\"ornstein
       compilation~\cite{HS91}. Compilation is on-going.
       \item[4]{\bf Proton-induced total reaction cross sections (10):}
       Proton-induced total reaction cross section data in
       Carlson's compilation~\cite{RFC96}.
       Compilation was completed by 2012 except for one article.
       \item[5]{\bf Nuclear resonance fluorescence (NRF) data (10):}
       Properties of resonances excited by $\gamma$-ray scattering and
       relevant to nondestructive assay (NDA) of fissile materials.
       All articles were compiled by 2012~\cite{SS11b}.
       \end{itemize}
       Similar checking was also done for other types of data
       (\eg
       data used in the IAEA Spallation Model Benchmarking~\cite{SL11},
       super-heavy elements production cross sections).
       
       Another new direction of extension is compilation of evaluated or
       recommended reaction data not distributed in
       the ENDF-6 format~\cite{MH09}.
       Initially such an attempt was made by NDS for the \EXFOR-VIEN
       (Various International Evaluated Neutron Data)
       file~\cite{KO84}.
       In 2012, compilation was done by NNDC and NDS for the thermal neutron data
       recommended
       by S.~Mughabghab~\cite{SM06} and Maxwellian averaged neutron capture
       cross sections at $kT$=30 keV recommended by
       Z.Y.~Bao et al.~\cite{ZYB00}.
       There are also similar attempts for charged-particle induced 
       isotope production cross sections (\eg~\cite{ST03}) and photoneutron
       reaction cross sections (\eg~\cite{VVV04}).
       
       Finally we would like to note that the completeness depends strongly on
       the range of the data types and availability.
       For example, data in conference proceedings, raw data in arbitrary unit,
       data not available from authors could be on the boundary of the scope.
       
       \section{QUALITY ASSURANCE}
       Quality assurance is another important issue for the \EXFOR library.
       The entire information of \EXFOR entries is mostly typed by \EXFOR
       compilers manually, and sometimes they have to type hundreds
       of numerical data lines not available in an electronic
       form.
       Even though the \EXFOR compiler of the originating center takes the greatest
       care during compilation,
       it is still impossible to eliminate all errors
       at the stage of compilation.
       On the other hand,
       \EXFOR users have more opportunity to compare
       different \EXFOR data sets 
       with their own experimental or theoretical data set for a certain range
       of reactions and quantities, and they are in a good position to detect
       errors.
       However, there was no well-established means of communication between
       \EXFOR users and NRDC.

       A turning point came when two valuable lists of suspicious \EXFOR
       entries (\eg a factor 1000 larger than usual values due to coding of
       barn instead of millibarn) were submitted by A.J.~Koning (NRG) and
       R.A.~Forrest (UKAEA) and discussed in the NRDC 2006 Meeting~\cite{OS06}.
       In order to improve the quality of the \EXFOR contents in collaboration
       between the \EXFOR users and NRDC, a new WPEC subgroup
       ``Quality Improvement of the \EXFOR Database (SG30)"~\cite{AK11} was
       coordinated by A.J.~Koning from 2007 to 2010, and detection of errors
       and their corrections were performed in a systematic manner.
       The initial important step was translation of contents of the \EXFOR
       database to the extended Computational Format (XC4) at NDS using
       the X4toC4 code~\cite{DEC01}.
       The detection of suspicious \EXFOR data sets were then mainly done by
       two methods: (1) detection of outliers by
       intercomparison of data points in XC4, and (2) comparison of
       data points in XC4 with prediction by TALYS~\cite{AK12}.
       The suspicious entries were further filtered by visual inspection
       using the JANIS display software~\cite{NS11} at NEA DB,
       and then checked against the original articles at NDS.
       Finally about 100 erroneous \EXFOR data sets were confirmed,
       and were corrected by the originating data centers.
       More details of these approaches are reported elsewhere~\cite{ED11}.
       
       The development of such systematic and semi-automatic detection is
       continuing at NEA DB (in collaboration with NRG)~\cite{XChk}
       involving data types not covered by the WPEC SG30 activity.
       Here are examples of additional inspections performed by NDS (with the number of detected
       erroneous data sets in parentheses):
       incident energy coded in MeV instead of in keV (29),
       level energies higher than 20 MeV or lower than 10 keV (59),
       reaction violating charge or mass conservation (17),
       partial data without specification of excitation level (288).
       
       Checking codes (ZCHEX, JANIS TRANS Checker) also support \EXFOR
       compilers to eliminate format and physical errors before submission
       of their \EXFOR entries.
       Various other inspections (\eg formatting, bibliographic
       information) are also being done regularly by NEA DB.
       All comments from \EXFOR users and data centers are registered to the
       \EXFOR Feedback List (http://www-nds.iaea.org/nrdc/error/),
       and the correction process is monitored by NDS.
       Digitization is also a key process to determine the quality of numerical
       data published in old articles.
       NDS has organised the IAEA Consultant Meeting
       on ``Benchmarking of Digitization Software" in 2012~\cite{NO13}
       to improve this process.

       \section{OTHER IMPROVEMENTS}
       Various other efforts are being made to
       improve the contents and accessibility of the \EXFOR library.
       One of the most important issues is the detailed documentation of
       uncertainties and covariances to support evaluation
       with minimum assumption.
       The error propagation described in articles is not satisfactory to
       provide enough information to evaluators in the most cases.
       Recently the \EXFOR format was extended to accommodate correlation 
       properties and covariance matrices in computer readable form, and
       guides were published to promote submission of detailed information
       by experimentalists~\cite{DLS12,WM13}.
       Archiving of time-of-flight spectra is also important when one needs to
       evaluate covariances between resonance parameters by error propagation
       from the primary measurable~\cite{BB12}.
       NDS is working for 
       compilation and documentation of time-of-flight spectra
       in collaboration with EC-JRC IRMM~\cite{PS12}.
       
       Another advance is seen in \EXFOR entries for prompt fission neutron
       spectra (PFNS).
       They are very rarely given in the absolute unit (\ie
       neutrons/energy/fission), and the coding method was not well
       standardised.
       Motivated by the currently on-going IAEA CRP on ``Prompt Fission Neutron
       Spectra of Actinides"~\cite{RCN13},
       all PFNS \EXFOR entries were upgraded thoroughly by data centers.
       In addition, PFNS for Pu, Am and Cm
       measured by Khlopin Radium Institute within the ISTC project were compiled by JAEA/NDC and NDS.
       Such improvements related to IAEA CRPs are also foreseen for data
       related to $\beta$-delayed neutron~\cite{DA12}
       and IRDFF library validation~\cite{RC12}.
       In order to improve accessibility to English translation of articles in
       Russian, systematic addition of English translation information to \EXFOR
       entries are on-going, led by CAJaD.
       
       Further improvement of formats is also discussed to make the contents of
       \EXFOR entries more understandable~\cite{RF13,DB12}.
       
       \section{CONCLUSIONS}
       The needs for experimental reaction data are always growing.
       Also more and more information in the \EXFOR library is expected to be
       machine readable according to development of various processing tools.
       NRDC is always trying to take various approaches to maintain \EXFOR as a
       very complete and error-free library.
       Feedback from \EXFOR users is also extremely important to
       achieve this goal.
       

       \end{document}